\documentclass[12pt]{article}
\usepackage{amsfonts}
\usepackage{amssymb}
\usepackage{amsbsy}

\input epsf.sty
\newcommand{\BEQ}{\begin{equation}}     
\newcommand{\BEA}{\begin{eqnarray}}
\newcommand{\EEQ}{\end{equation}}       
\newcommand{\EEA}{\end{eqnarray}}
\newcommand{\eps}{\varepsilon}          
\newcommand{\del}{\delta}

\newcommand{\g}{{\mathfrak{g}}}
\newcommand{\so}{{\mathfrak{so}}}
\newcommand{\h}{{\mathfrak{h}}}

\newcommand{\p}{{\mathfrak{p}}}
\newcommand{\sv}{{\mathfrak{sv}}}

\newcommand{\slin}{{\mathfrak{sl}}}

\newcommand{\alt}{{\mathfrak{alt}_1}}
\newcommand{\sch}{{\mathfrak{sch}}}
\newcommand{\conf}{{\mathfrak{conf}}}

\newcommand{\al}{\alpha}

\newcommand{\half}{{1\over 2}}

\newcommand{\R}{\mathbb{R}}
\newcommand{\C}{\mathbb{C}}
\newcommand{\Z}{\mathbb{Z}}
\newcommand{\D}{{\rm d}}                
\newcommand{\II}{{\rm i}}               
\newcommand{\wit}[1]{\widetilde{#1}}    
\newcommand{\rar}{\rightarrow}          
\renewcommand{\vec}[1]{\boldsymbol{#1}} 

\catcode`\@=11
\def\numberbysection{\@addtoreset{equation}{section}
        \def\theequation{\thesection.\arabic{equation}}}
\numberbysection

\def\@xthm#1#2{\@begintheorem{#2}{\csname the#1\endcsname}{}\ignorespaces}
\def\@ythm#1#2[#3]{\@opargbegintheorem{#2}{\csname
       the#1\endcsname}{#3}\ignorespaces}%

\def\@begintheorem#1#2#3{\par\addvspace{8pt plus3pt minus2pt}%
              \noindent{\csname#1headfont\endcsname#1\ \ignorespaces#3 #2.}%
              \csname#1font\endcsname\hskip.5em\ignorespaces}
\def\@endtheorem{\par\addvspace{8pt plus3pt minus2pt}\@endparenv}

\newtheorem{lemma}{Lemma}[section]
\newtheorem{proposition}{Proposition}[section]

\newtheorem{example}{Example}[section]
\newtheorem{remark}{Remark}[section]

\newcommand{\qed}{q.e.d.}

\begin{document}

\parskip 2mm 
\vskip 1.5 cm

\begin{titlepage} 
\begin{center}
{\Large \bf The Poincar\'e algebra in the context of ageing systems: \\
Lie structure, representations, Appell systems \\ 
and coherent states}
\end{center}
\vskip 2.0 cm
\begin{center}
{\bf Malte Henkel}$^{a}$, {\bf Ren\'e Schott}$^{b,c}$,  
{\bf Stoimen Stoimenov}$^{d}$ and {\bf J\'er\'emie Unterberger}$^{b}$
\end{center}
\vskip 0.5 cm
\begin{center} 
$^a$Groupe de Physique Statistique, Institut Jean Lamour (CNRS UMR 7198),\\
Universit\'e de Lorraine Nancy,  
B.P. 70239, F -- 54506 Vand{\oe}uvre-l\`es-Nancy Cedex, France \\
$^b$Institut \'Elie Cartan (IECN -- CNRS UMR 7502), 
Universit\'e de Lorraine Nancy, \\ 
B.P. 70239, F -- 54506 Vand{\oe}uvre-l\`es-Nancy Cedex, France \\
$^c$LORIA (CNRS UMR 7503), Universit\'e de Lorraine Nancy, \\ 
B.P. 70239, F -- 54506 Vand{\oe}uvre-l\`es-Nancy Cedex, France \\
$^d$Institute of Nuclear Research and Nuclear Energy,
Bulgarian Academy of Sciences, \\
BG - 1784 Sofia, Bulgaria
\end{center}

\begin{abstract}
By introducing an unconventional realization of the Poincar\'e algebra $\mathfrak{alt}_1$
of special relativity as conformal transformations, 
we show how it may occur as a dynamical symmetry algebra for ageing
systems in non-equilibrium statistical physics and give some applications, 
such as the computation of two-time correlators. 
We also discuss infinite-dimensional extensions of $\mathfrak{alt}_1$ in this
setting. Finally, we construct canonical Appell systems, 
coherent states and Leibniz function for $\alt$ as a tool for bosonic quantization.
\end{abstract}

\end{titlepage}

\section{Introduction}	
Ageing phenomena occur widely in physics: glasses, 
granular systems or phase-ordering kinetics are just a few examples, see e.g.
\cite{Bray94,Cugl02,Henk06,Henk10} for reviews.
Ageing phenomena may typically arise in the presence of two competing,
globally equivalent, steady-states: except for very particular initial
preparations, the physical system does not relax to any of these; rather
there appear ordered domains which grow in size with
time and which are separated by fluctuating boundaries.\footnote{A typical
experimental method to achieve this is to prepare a system in a disordered
`high-temperature' initial state and then to `quench' it by lowering very
rapidly the temperature below the critical temperature $T_c$
such that ergodicity is broken and several distinct steady-states appear.} 
A convenient way to describe this sort of system is
through a Langevin equation, which might schematically be written as
\BEQ \label{lang}
2{\cal M} \partial_t \phi = - \frac{\delta {\cal F}}{\delta \phi} + \eta
\EEQ
where $\phi=\phi(t,\vec{r})$ stands for the physical order-parameter
(here assumed to be non-conserved),
$\cal F$ is the Ginsburg-Landau functional, the `mass' $\cal M$ plays the
r\^ole of a kinetic coefficient and $\eta$ 
describes a gaussian, delta-correlated noise.
While it is well-accepted \cite{Bray94}
that systems of this kind should display some sort of dynamical scaling when
brought into the situation sketched above, the question has been raised whether
their non-equilibrium dynamics might possess more  symmetries than merely
scale-invariance \cite{Henk02}. At first sight, the noisy terms in the
Langevin equation (\ref{lang}) might appear to exclude any non-trivial answer.
However, a more refined answer is possible. One may consider (\ref{lang}) as
the classical equation of motion of an associated field-theory, whose action
reads
\BEQ
S[\phi,\wit{\phi}] = S_0[\phi,\wit{\phi}] + S_b[\wit{\phi}]
\EEQ
where $\wit{\phi}$ is the response-field associated to the order-parameter
field $\phi$. Here, the `noise' as described by the random force $\eta$ only
enters into the second term $S_b[\wit{\phi}]$. In many cases, the so-called
`noise-less' part $S_0$ takes a free-field form
\BEQ
S_0[\phi,\wit{\phi}] = \int\!\D t\D\vec{r}\: \wit{\phi}
\left( 2{\cal M}\partial_t - \Delta \right) \phi
\EEQ
which has the important property of being Galilei-invariant. If that is the
case, the Bargman superselection rules coming from the Galilei-invariance of
$S_0$ allow to show that {\em all} $n$-point correlation and response functions
of the theory can be expressed in terms of certain $(n+2)$-point correlation
function of an effective deterministic theory whose action is simply $S_0$
\cite{Pico04}. This result does not depend on $S_0$ being a free-field action but
merely on its Galilei-invariance \cite{Baum05}. In order to study the
properties of the stochastic Langevin equation (\ref{lang}), it is hence
sufficient to concentrate on the properties of its deterministic part (where
$\eta$ is dropped) which reduces the problem to the study of those dynamical
symmetries of non-linear partial differential equations which extend dynamical
scaling. For a systematic exposition, with many explicit tests, see \cite{Henk10}.
The symmetry properties of the deterministic 
part of this kind of problems will be studied in this paper.

\begin{figure}[t]
\centerline{\epsfxsize=1.1in\ \epsfbox{
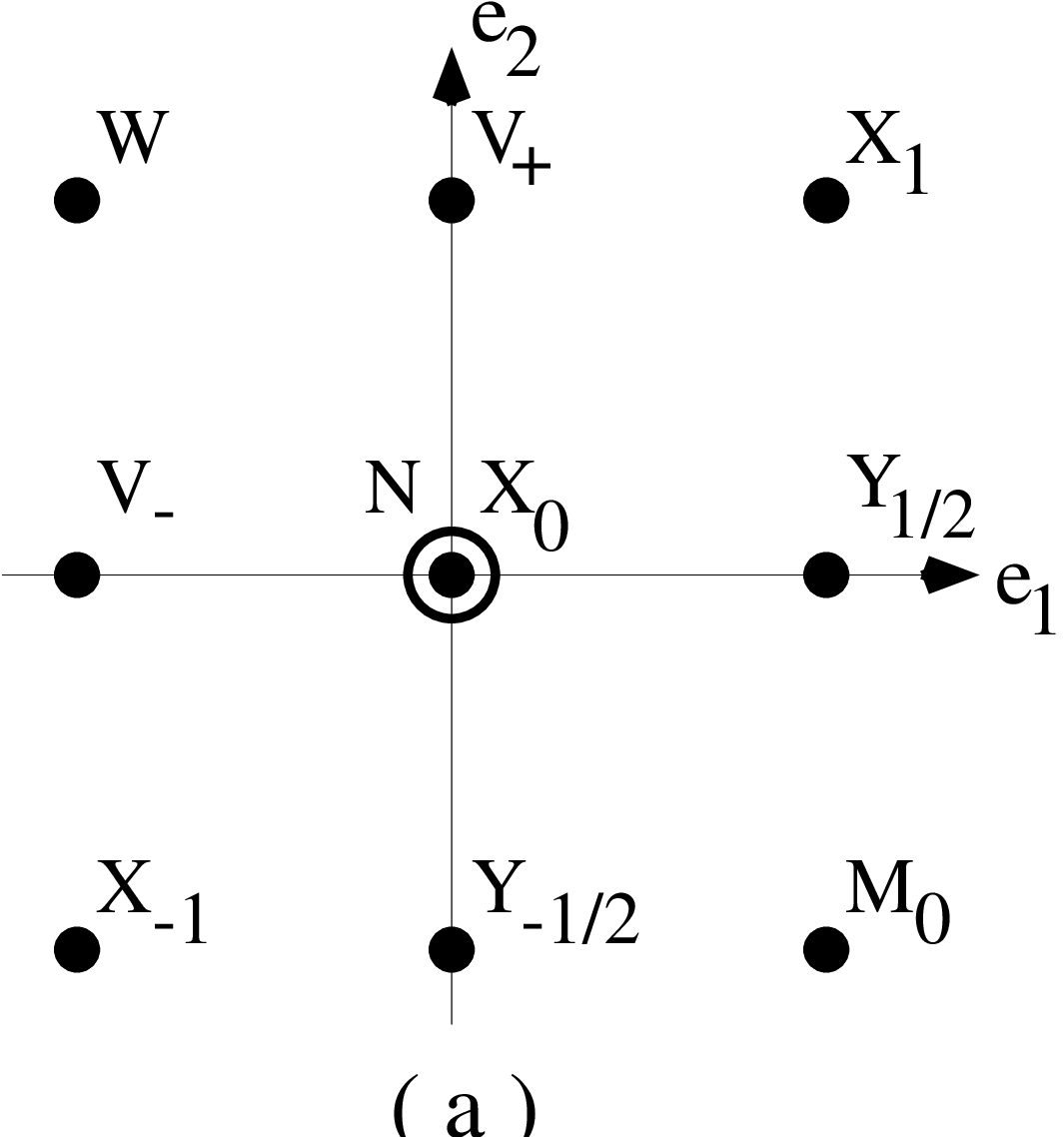} ~
\epsfxsize=1.1in\epsfbox{
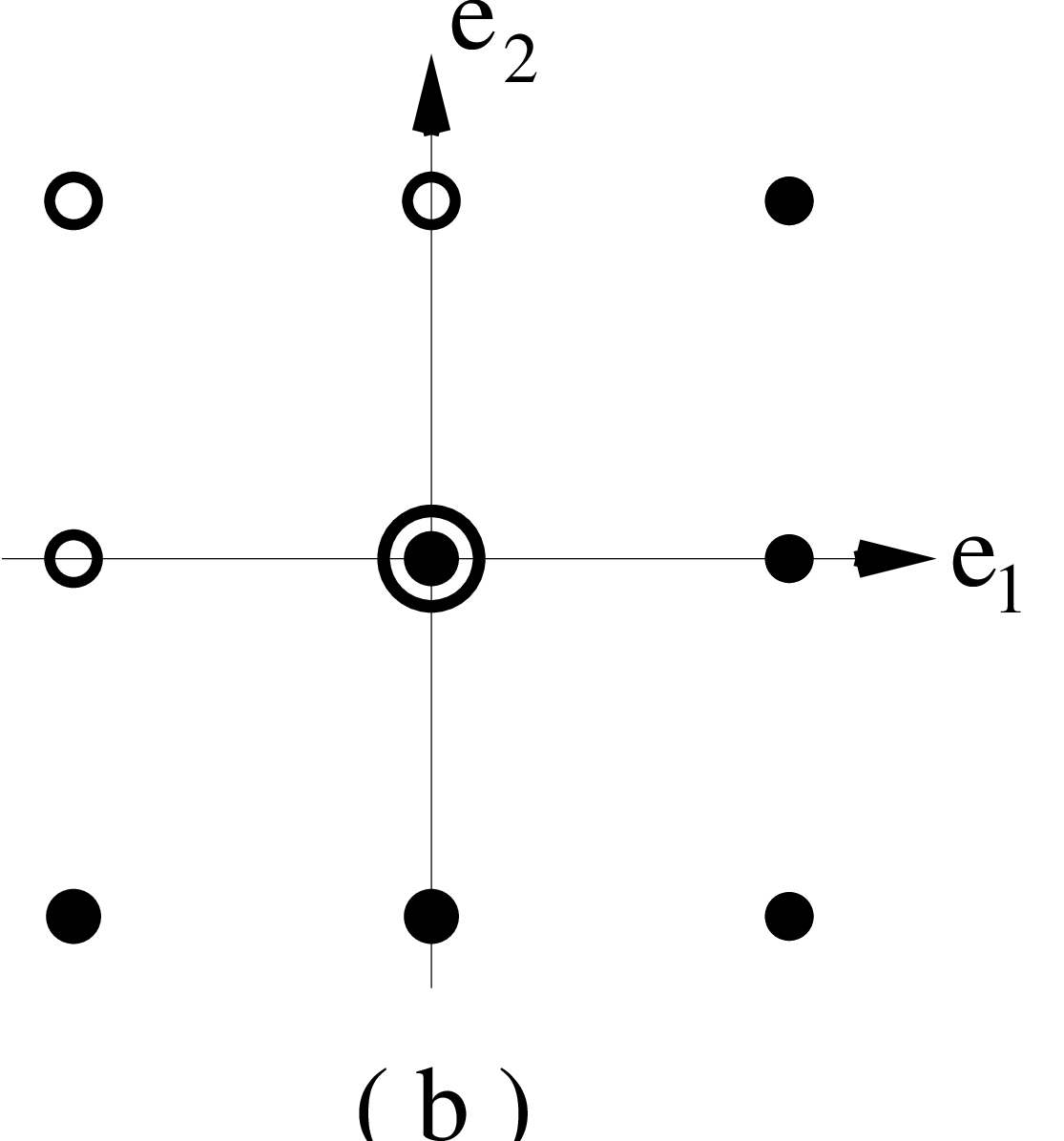} ~
\epsfxsize=1.1in\epsfbox{
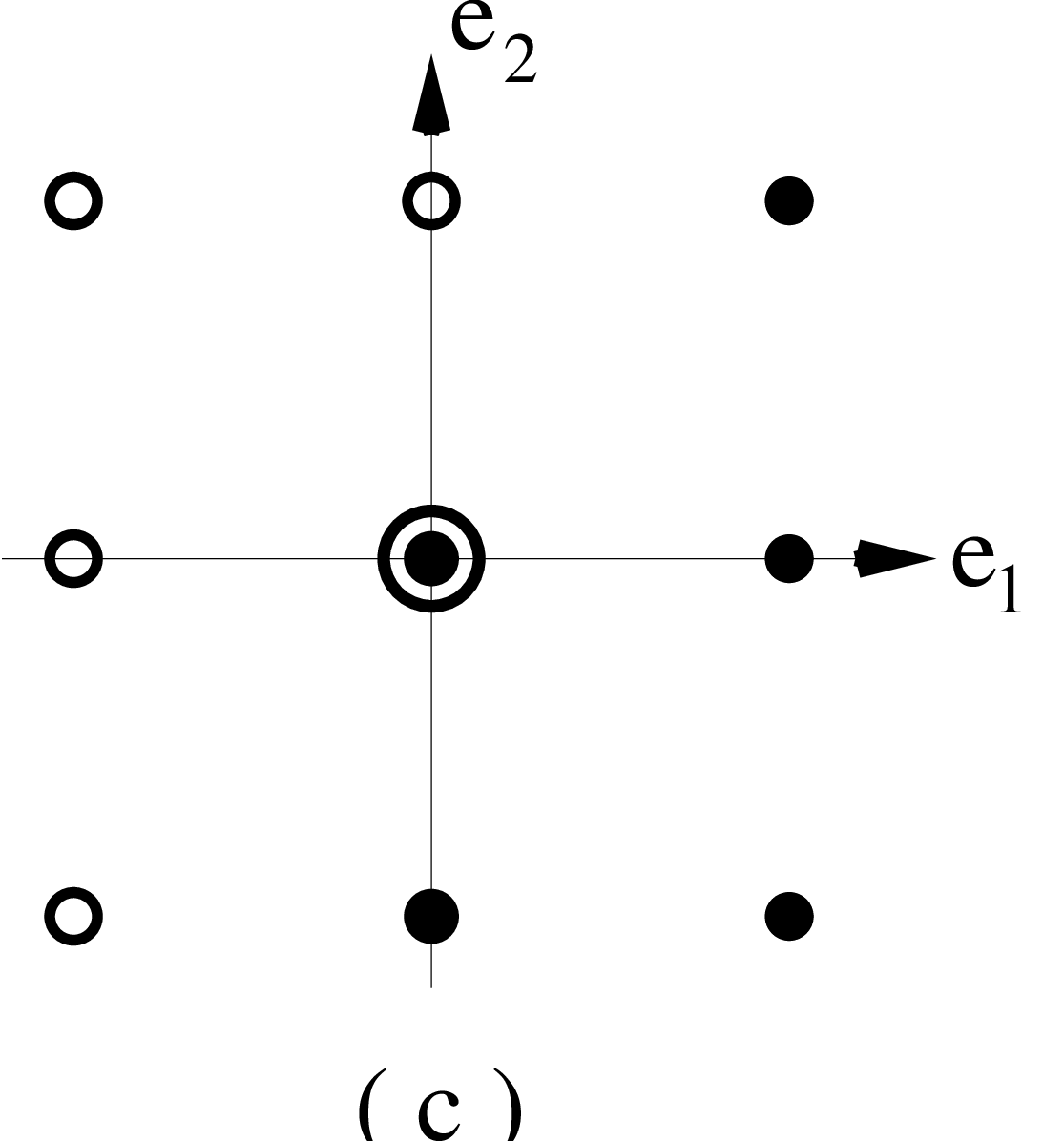} ~
\epsfxsize=1.1in\epsfbox{
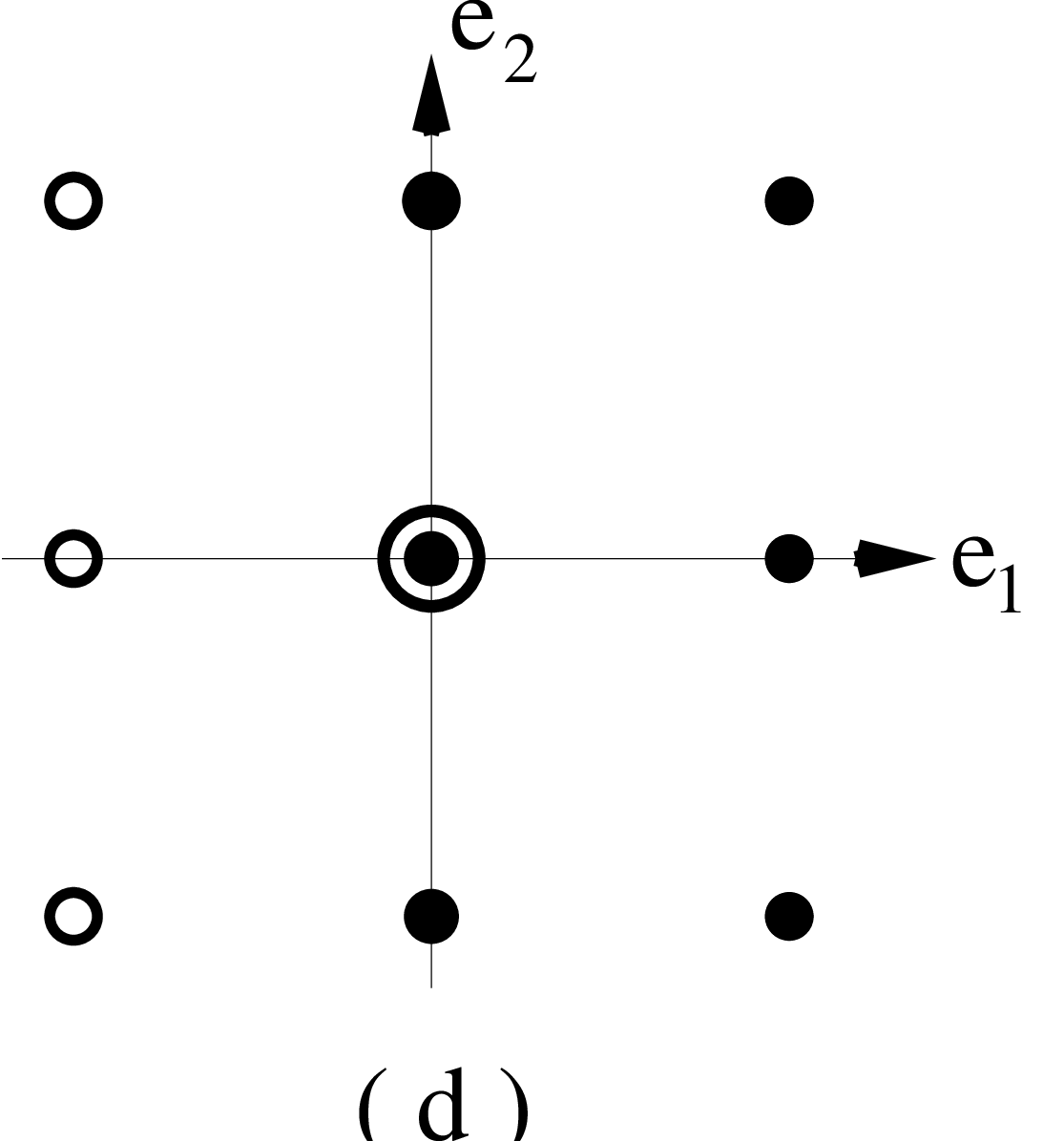}
}
\caption[Root space]{\small
(a) Root diagram of the complex Lie algebra $B_2$ and the
identification of the generators (\ref{eq:zetaGen})
of the complexified conformal Lie algebra
$(\mathfrak{conf}_3)_{\mathbb{C}}\supset(\mathfrak{sch}_1)_{\mathbb{C}}$.
The double circle in the centre denotes the Cartan subalgebra.
The generators belonging to the three non-isomorphic parabolic subalgebras
\protect{\cite{Henk03}}
are indicated by the full points, namely
(b) $\wit{\mathfrak{sch}}_1$, (c) $\wit{\mathfrak{age}}_1$ and
(d) $\wit{\mathfrak{alt}}_1$.
\label{Bild1}}
\end{figure}

In the context of phase-ordering kinetics, the Schr\"odinger algebra $\sch_1$
\cite{Nied72} has played an important r\^ole.\footnote{While in the
literature it is usually stated that the Lie algebra
$\mathfrak{sch}_1$ was first written down by Lie \cite{Lie1881},
its elements already occur almost forty years earlier in Jacobi's lectures on
analytical mechanics \cite{Jacobi1843}.} In what follows we shall
restrict to one space dimension and we recall in Figure~\ref{Bild1} through a
root diagram the definition of $\widetilde{\sch}_1$ as a parabolic subalgebra
of the conformal algebra $\conf_3$ \cite{Burd73,Henk03}. The inclusion
$\sch_1\subset (\conf_3)_{\C}$ can be  realized by considering the `mass'
$\cal M$ as an additional coordinate. It is
convenient to perform a Fourier-Laplace transform with respect to $\cal M$,
with the dual coordinate $\zeta$. Then the generators of
$(\conf_3)_{\C}$ read explicitly
\BEA
X_{-1} &=& -\partial_t \;\; , \;\; Y_{-{1\over 2}} = -\partial_r, \quad M_0 
= \II\partial_{\zeta}, \quad
Y_{1\over 2} = -t\partial_r+\II r\partial_{\zeta }\nonumber\\
X_0 &=& -t\partial_t-{1\over 2}r\partial_r-{x\over 2}, \quad 
X_1 = -t^2\partial_t-tr\partial_r+{\II\over 2}r^2\partial_{\zeta}-xt
\nonumber \\
N &=& -t\partial_t+\zeta\partial_{\zeta}, \quad 
W = -\zeta^2\partial_{\zeta}-\zeta r\partial_r
-{1\over 2}r^2\partial_t-x\zeta\nonumber\\
V_- &=&  -\zeta \partial_r-r\partial_t, \quad  
V_+ = -2tr\partial_t-2\zeta r\partial_{\zeta }-
(r^2+2\II\zeta t)\partial_r-2xr.
\label{eq:zetaGen}
\EEA
and the correspondence with the root vectors is illustrated in
figure~\ref{Bild1}a.

The complete list of non-isomorphic
parabolic subalgebras of $(\mathfrak{conf}_3)_{\mathbb{C}}$ is as follows \cite{Henk03}
\BEA
\wit{\mathfrak{sch}}_1 &=& \langle X_{-1,0,1}, Y_{-\frac{1}{2},\frac{1}{2}}, M_0, N \rangle
\nonumber \\
\wit{\mathfrak{age}}_1 &=& \langle X_{0,1}, Y_{-{1\over 2},{1\over 2}}, M_0, N
\rangle
\\
\wit{\mathfrak{alt}}_1 &=& \langle
D,X_1,Y_{-{1\over 2},{1\over 2}},M_0,N,V_+\rangle
\nonumber
\EEA
and these definitions are illustrated in figure~\ref{Bild1}bcd.
Here we used the generator $D$ of the full dilatations\footnote{In physics, if one considers
the space-time rescaling $t\mapsto \lambda^z t$, 
$\vec{r} \mapsto \lambda \vec{r}$ with constant
$\lambda$, the quantity $z$ is called {\em dynamical exponent}. 
Integrating the infinitesimal dilatation generator $X_0$ in (\ref{eq:zetaGen}), 
one finds $z=2$, whereas the dilatation generator $D$ in (\ref{eq:ddef})
gives $z=1$.} 
\BEQ\label{eq:ddef}
D := 2 X_0 - N = -t\partial_t-r\partial_r-\zeta \partial_{\zeta }-x
\EEQ
For applications to non-equilibrium physics, it is of interest to
consider as well the corresponding {\it ``almost-parabolic subalgebras''}
without the generator $N$ \cite{Stoi05}, namely
\BEA
\mathfrak{sch}_1 &=& \langle X_{-1,0,1},Y_{\pm\frac{1}{2}}, M_0\rangle
\nonumber \\
\mathfrak{age}_1 &=& \langle X_{0,1}, Y_{\pm{1\over 2}}, M_0\rangle
\\
\mathfrak{alt}_1 &=& \langle D,X_1,Y_{\pm{1\over 2}},M_0,V_+\rangle
\nonumber
\EEA
We shall show in Section 2.1 that $\mathfrak{alt}_1$ is isomorphic to the
Poincar\'e algebra $\mathfrak{p}_3$ (well-known from relativistic field-theory)
through a non-conventional realization of the latter;
a correspondence which at first thought might appear surprising.

While the explicit representation (\ref{eq:zetaGen}) concerns the linear
free Schr\"odinger equation ${\cal S}\phi=0$, with
${\cal S} = 2 M_0 X_{-1} - Y_{-1/2}^2$, Schr\"odinger-invariance
can also be proven for s\'emi-linear Schr\"odinger equations of the form
\BEQ \label{NLSE}
{\cal S}\phi = g F(\phi,\wit{\phi})
\EEQ
where $g$ is a dimensionful coupling constant, hence it transforms
under the action of scaling or conformal transformations. The corresponding
representations have been explicitly derived in the case of a variable
mass for $(\conf_3)_{\C}$ \cite{Stoi05} and its subalgebras and for a fixed
mass for $\sch_1$ and for $\mathfrak{age}_1$ \cite{Baum05},
{}from which the form of the
potential $F$ in (\ref{NLSE}) can be deduced. Supersymmetric extensions
of the Schr\"odinger algebra are discussed in \cite{Henk05}.

While these examples and others already illustrate the intensive study of the
Schr\"odinger algebra and of its subalgebras \cite{Boye76},
the other non-trivial subalgebra ${\mathfrak{alt}}_1$ has so far received much
less attention.\footnote{This algebra had been identified first in \cite{Havas78}, 
under the name of `{\em conformal galilean algebra}' $\mbox{\sc cga}(1)$. 
In recent years, especially string theorists have pursued the
study of its representations, often in the context of variants of the AdS/CFT correspondence.
It can be shown that $\mathfrak{sch}_1$ and $\mathfrak{alt}_1$ are essentially the
only possible distinct non-relativistic limits of the conformal algebra, for light-like and
time-like geodesics, respectively \cite{Duval09}.}
One of the few results established so far concerns the
non-relativistic limit of the conformal algebra
$(\mathfrak{conf}_3)_{\mathbb{C}}$. For a dynamical mass, that is the
dynamical symmetry algebra of the massive Klein-Gordon equation
\BEQ
\left( \frac{1}{c^2}\frac{\partial^2}{\partial t^2} + \frac{\partial}{\partial\vec{r}}\cdot\frac{\partial}{\partial\vec{r}}
-{\cal M}^2 c^2\right) \phi_{\cal M}(t,\vec{r})=0
\EEQ
and the non-relativistic limit is obtained by letting the speed of light
$c\to\infty$. Contrary to widely held beliefs (which go back at least
to \cite{Baru73}), it turned out that in this limit
$(\mathfrak{conf}_3)_{\mathbb{C}}\to \mathfrak{alt}_1 \not\cong\mathfrak{sch}_1$ 
and furthermore this is {\em not}
a group contraction \cite{Henk03}.
In this paper, we study the  Lie structure of $\alt$,
give matrix- as well as dual representations, 

characterize Appell systems in connection with coherent states and Leibniz function.
The organization of the paper is as follows: 
Section~2 concerns the Lie structure of $\mathfrak{alt}_1$
where in particular we study infinite-dimensional 
extensions and also discuss applications to the
computation of covariant two-point functions. 
Casimir operators and matrix representations are provided in Section~3.
Section~4 focuses on Cartan decomposition and dual representations.
A smooth introduction to Wick products and Appell polynomials is given in Section~5.
Appell systems of $\mathfrak{alt}_1$ are characterized in Section~6.
Calculations concerning coherent states 
and Leibniz function are contained in Section~7 and we conclude in Section~8.
Some of our results were announced earlier \cite{HSSU}. 

\section{A Brief Perspective on the Algebra $\mathfrak{alt}_1$}

We shall  take in this section a closer look at the abstract Lie
algebra $\mathfrak{alt}_1$ and its representations; 
we shall also see that, like the algebra $\sch_1$, it can be embedded naturally
in an infinite-dimensional Lie algebra ${\cal W}$ which is an
extension of the algebra $\mathrm{Vect}(S^1)$ of vector fields on the circle. 
Quite strikingly, we shall find on our way a
'no-go theorem' that proves the impossibility of a conventional extension of the embedding
$\mathfrak{alt}_1\subset\conf_3$ on the one hand, and a surprisingly
simple geometric interpretation of
${\cal W}$ that hints at a possible connection with $\sv$.

\subsection{The abstract Lie algebra $\mathfrak{alt}_1$}

Elementary computations make it clear that
\BEQ
\mathfrak{alt}_1=\langle V_+,D,Y_{-\half}\rangle\ltimes \langle X_1,Y_{\half},M_0\rangle
=: \g\ltimes\h
\EEQ
is a semi-direct product of $\g\cong \slin(2,\R)$ 
by a three-dimensional commutative Lie algebra
$\h$; the vector space $\h$ is the irreducible spin-1 
real representation of $\slin(2,\R)$, which
can be identified with $\slin(2,\R)$ itself with the adjoint action.
So one has the following

\begin{proposition} The following Lie algebra isomorphisms
hold true. First,
\BEQ
\mathfrak{alt}_1 \cong \slin(2,\R)\otimes\R[\eps]/\eps^2,
\EEQ
 where $\eps$ is a 'Grassmann' variable. Second,
\BEQ
\mathfrak{alt}_1 \cong \p_{3}
\EEQ
 where $\p_3 \cong \so(2,1)\ltimes\R^3$ is the relativistic Poincar\'e 
 algebra in (2+1)-dimensions.
\end{proposition}

\noindent {\bf Proof:} 
We shall establish the first isomorphism explicitly. Take a basis
$(L_1,L_0,L_{-1})$ of $\slin(2,\R)$ such that
\begin{displaymath}
[L_0,L_1]=-L_1,\ [L_0,L_{-1}]=L_{-1},\ [L_1,L_{-1}]=2L_0
\end{displaymath}
These generators may be written in terms of the anticommuting
Pauli matrices $(\sigma_x,\sigma_y,\sigma_z)$ as follows
$L_1=(\sigma_x+\II\sigma_y)/2,\ L_0=-\sigma_z/2$ and 
$L_{-1}=-(\sigma_x-\II\sigma_y)/2$. Then let
\begin{displaymath}
L_i^{\eps}:=L_i\otimes{\eps}
\end{displaymath}
($i=-1,0,1$) where $\eps$ is a Grassmann variable. Then the linear map $\Phi:\mathfrak{alt}_1\to\slin(2,\R)\otimes\R[\eps]/\eps^2$ defined by
\BEA
\Phi(V_+) &=&L_1 \;\; , \;\;  \Phi(D)\:=\:L_0 \;\; , \;\;  
\Phi(Y_{-\half})\:=\:L_{-1} \nonumber \\
\Phi(X_1) &=&\half L_1^{\eps} \;\; , \;\;
\Phi(Y_{\half})\:=\:L_0^{\eps} \;\; , \;\;  \Phi(M_0)\:=\:L_{-1}^{\eps}
\nonumber
\EEA
is a Lie isomorphism.

The second relation is obvious from the Lie isomorphism 
$\so(2,1)\cong\slin(2,\R)$. \hfill \qed

In particular, the representations of $\mathfrak{alt}_1 \cong\p_3$
are  well-known since Wigner studied them in the 30'es.

\subsection{Central extensions: an introduction}
Consider any Lie algebra $\g$ and an 
antisymmetric real two-form $\alpha$ on $\g$.
Suppose that its Lie bracket $[\ , ]$ can be 'deformed' into a new
Lie bracket $\widetilde{[\ , ]}$ on 
$\tilde{\g}:=\g\times \R K$, where $[K,\g]=0$,
by putting $\widetilde{[(X,0),(Y,0)]}=([X,Y],\alpha(X,Y))$. 
Then $\tilde{\g}$ is called a
{\it central extension} of $\g$. The Jacobi identity is equivalent with
the nullity of the totally antisymmetric three-form
$\D\al:\Lambda^3(\g)\to\R$ defined by
\BEQ
\D\alpha(X,Y,Z)=\al([X,Y],Z)+\al([Y,Z],X)+\al([Z,X],Y).
\EEQ
Now we say that two central extensions $\g_1,\g_2$ of $\g$ 
defined by $\al_1,\al_2$ are equivalent if
$\g_2$ can be obtained from $\g_1$ by substituting 
$(X,c)\mapsto (X,c+\lambda(X))$ $(X\in\g)$ for
a certain 1-form $\lambda\in\g^*$, that is, by changing 
the non-intrinsic embedding of $\g$
into $\tilde{\g}_1$. In other words, $\al_1$ and $\al_2$ are equivalent if
$\al_2-\al_1=\D\lambda$, where $\D\lambda(X,Y)=\langle \lambda,[X,Y]\rangle.$
The operator $\D$ can be made into the differential of a complex 
(called Chevalley-Eilenberg complex),
and the preceding considerations make it clear that the classes of
equivalence of central extensions of $\g$ make up a vector space
$H^2(\g)=Z^2(\g)/B^2(\g)$, where $Z^2$ is
the space of {\it cocycles} $\alpha\in\Lambda^2(\g^*)$ verifying $\D\alpha=0 $,
and $B^2$ is the space of {\it coboundaries} $\D\lambda$, $\lambda\in\g^*$.

Let us see how this applies to $\mathfrak{alt}_1$.

\begin{proposition}
 The Lie algebra $\mathfrak{alt}_1$ has no non-trivial central extension:
$H^2(\alt)=0$.
\end{proposition}

\noindent {\bf Proof:} 
Of course, this is a consequence of the fact that 
Poincar\'e algebras have no non-trivial central extensions,
but let us give a proof in this simple example to see how computations work. 
Note that ad $L_0$ acts diagonally on the
generators $(L_i)$ and $(L_i^{\eps})$ $(i=-1,0,1)$, 
defining a ${-1,0,1}$-valued graduation
$\delta$ on $\mathfrak{alt}_1$. It is then well-known that 
$\al$ is cohomologous to a cocycle $\al'$ such
that $\al'(Z_i,Z_j)=0$ if $Z_i,Z_j$ are homogeneous 
generators such that $\del(Z_i)+\del(Z_j)\not=0$
(see for instance \cite{GuiRog05}, chapter 4), 
so we may just as well assume this is already the case for $\al$.
Then $\al$ is defined by
\begin{displaymath}
a:=\al(L_1,L_{-1}),\ a^{\eps}:=\al(L_1^{\eps},L_{-1}^{\eps}),\ b:=\al(L_1,L_{-1}^{\eps}),\
b^{\eps}:=\al(L_1^{\eps},L_{-1}),\ c:=\al(L_0,L_0^{\eps}).
\end{displaymath}
The non-trivial Jacobi identities
\BEA
0 &=&\D\al(L_1,L_{-1},L_0^{\eps})\:=\:2c+b-b^{\eps} \;\; , \;\; 
0\:=\:\D\al(L_1^{\eps},L_{-1},L_0^{\eps})\:=\: a^{\eps},
\nonumber \\
0 &=& \D\alpha(L_1,L_{-1}^{\eps},L_0^{\eps})\:=\:a^{\eps} \;\; , \;\;
\hspace{1.5truecm}
0\:=\: \D\al(L_1^{\eps},L_{-1},L_0)\:=\: -2c
\nonumber
\EEA
give $c=a^{\eps}=0$ and $b=b^{\eps}$. But the central extension 
$\al$ is then trivial: it is killed by substituting
$L_0\to L_0+2aK, L_0^{\eps}\to L_0^{\eps}+2bK.$ \hfill \qed

A new situation arises upon embedding $\mathfrak{alt}_1$ into an
infinite-dimensional Lie algebra.
\begin{remark} In $d=2$ spatial dimensions, 
a so-called `exotic' central extension exists for the algebra 
$\mathfrak{alt}_2$ \cite{Luki06}.
\end{remark}

\subsection{Infinite-dimensional extension of $\mathfrak{alt}_1$}

The Lie algebra $\mathrm{Vect}(S^1)$
of vector fields on the circle has a long story
in mathematical physics. It was discovered by Virasoro in 1970
\cite{Vira70,Clav71} that $\mathrm{Vect}(S^1)$ has a
one-parameter family of central extensions which yield the so-called
{\em Virasoro algebra}
\BEQ
{\mathfrak{vir}} := \mathrm{Vect}(S^1)\oplus\R K
=\langle (L_n)_{n\in\Z},K\rangle
\EEQ
with Lie brackets ($c\in\R$ is a parameter and is called the
{\em central charge})
\BEQ
[K,L_n]=0, \quad [L_n,L_m]=(n-m)L_{n+m}
+\delta_{n+m,0} \frac{c}{12} n(n^2-1)K\quad
\EEQ
When $c=0$, one retrieves $\mathrm{Vect}(S^1)$ by
identifying the $(L_n)$ with the usual Fourier
basis $(e^{\II n\theta}\D\theta)_{n\in\Z}$ of periodic vector fields on
$[0,2\pi]$, or with $L_n \mapsto \ell_n := -z^{n+1}{\D\over \D z}$ with
$z:=e^{\II\theta}$. Note in particular that $\langle L_{-1},L_0,L_1\rangle$
is isomorphic to $\slin(2,\R)$, with Lie brackets given in subsection 2.1,
and that the Virasoro cocycle restricted to $\slin(2,\R)$ is $0$,
as should be (since $\slin(2,\R)$ has no non-trivial central extensions).

The Schr\"odinger algebra $\sch_1$ can be embedded
into the infinite-dimensional Lie algebra $\sv$ (introduced in 1994
\cite{Henk94}) which is spanned by the generators $\langle L_n,Y_m,M_n\rangle$,
with non-vanishing commutators
\BEA
[L_n,L_p] &=& (n-p)L_{n+p} \;\; , \;\;
[L_n,Y_m] \:=\: (n/2 -m)Y_{n+m} \nonumber\\
{} [L_n,M_p] &=& -pM_{n+p}\;\;, \;\; [Y_m,Y_{m'}] \:=\: (m-m')M_{m+m'}
\EEA
with $n,p\in\Z$ and $m,m'\in\Z+\half$.  Note that $\sv$ is a
semi-direct product of
$\mathrm{Vect}(S^1)$ with an infinite-dimensional nilpotent Lie algebra.
Its mathematical structure is analyzed in detail in \cite{RogUnt05,Unte11} and
supersymmetric extensions are
discussed in \cite{Henk05}. There is only one class of central extensions of
$\sv$, given by the extension by zero of the Virasoro cocycle \cite{Henk94}.

An analogous embedding holds for $\alt$, namely
$\mathfrak{alt}_1\cong \slin(2,\R)\otimes\R[\eps]/\eps^2$ can be
embedded into the Lie algebra
\BEQ
{\cal W}:=\mathrm{Vect}(S^1)\otimes\R[\eps]/\eps^2
=\langle L_n\rangle_{n\in\Z}\ltimes \langle L_n^{\eps}
\rangle_{n\in\Z},
\EEQ
with Lie bracket
\BEQ
[L_n,L_m]=(n-m)L_{n+m},\ [L_n,L_m^{\eps}]
=(n-m)L_{n+m}^{\eps},\ [L_n^{\eps},L_m^{\eps}]=0. \label{gl:W}
\EEQ
These brackets come out naturally when one puts 
$\cal W$ in the $2\times 2$-matrix form
\BEQ
L_n\mapsto \left( \begin{array}{cc} \ell_n & 0\\ 0& \ell_n\end{array}\right),
\ L_n^{\eps}\mapsto \left(\begin{array}{cc} 0& \ell_n\\ 0& 0 \end{array}\right)
\EEQ
leading to straightforward generalizations
(see in particular \cite{RogUnt05,Unte11} for a deformation of $\sv$ that can be
represented as upper-triangular $3\times 3$ Virasoro matrices instead).

In terms of the standard representations of $\mathrm{Vect}(S^1)$
as  modules of $\al$-densities
${\cal F}_{\al}=\{ u(z)(\D z)^{\al}\}$ with the action
\BEQ
f(z){\D\over \D z} \left(u(z)(\D z)^{\al}\right)
=(fu'+\al f' u)(z) (\D z)^{\al},
\EEQ
we have

\begin{proposition}
\BEQ
{\cal W}\cong {\mathrm{Vect}}(S^1)\ltimes {\cal F}_{-1}
\EEQ
\end{proposition}

\noindent {\bf Proof:} 
Immediate from the obvious isomorphism of ${\mathrm{Vect}}(S^1)$ 
(with the adjoint action) with the ${\mathrm{Vect}}(S^1)$-module ${\cal F}_{-1}$.
\hfill \qed

It can be easily shown that $\cal W$ has {\it two} linearly independent central extensions:
\begin{enumerate}
\item the natural extension to $\cal W$ of the Virasoro cocycle on
$\mathrm{Vect}(S^1)$, namely $[\ ,\ ]=\widetilde{[\ ,\ ]}$ 
except for $\widetilde{[L_n,L_{-n}]}=n(n^2-1)K+2nL_0.$
In other words, $\mathrm{Vect}(S^1)$
is centrally extended, but its action on
${\cal F}_{-1}$ remains unchanged.
\item the cocycle $\omega$ which is zero on $\Lambda^2(\mathrm{Vect}(S^1))$ and
$\Lambda^2({\cal F}_{-1})$, and defined on $\mathrm{Vect}(S^1)\times {\cal F}_{-1}$ by
\BEQ
\omega(L_n,L_m^{\eps})=\delta_{n+m,0}\, n(n^2-1)K^{\eps}
\EEQ
\end{enumerate}
The independence of these two central charges is nicely illustrated through the
following example: consider the generators $V_n$ and $V_n'$ ($n\in\Z$) of
two commuting Virasoro algebras with central charges $c$ and $c'$. Then
identify
\BEQ
L_n\mapsto \left( \begin{array}{cc} V_n+V_n' & 0\\ 0& V_n+V_n'
\end{array}\right),
\ L_n^{\eps}\mapsto \left(\begin{array}{cc} 0& V_n\\ 0& 0 \end{array}\right),
\ K \mapsto \left(\begin{array}{cc} 1 & 0 \\ 0 & 1 \end{array}\right),
\ K^{\eps} \mapsto \left(\begin{array}{cc} 0 & 1 \\ 0 & 0 \end{array}\right)
\EEQ
and the non-vanishing commutators become
\BEA
{}[L_n,L_m] &=& (n-m)L_{n+m} +\frac{c+c'}{12}\left(n^3-n\right)\delta_{n+m,0} K,\
\nonumber \\
{}[L_n,L_m^{\eps}] &=& (n-m)L_{n+m}^{\eps} 
+ \frac{c}{12}\left(n^3-n\right)\delta_{n+m,0} K^{\eps}
\EEA

A natural related question is: can one deform the extension of $\mathrm{Vect}(S^1)$ by the
$\mathrm{Vect}(S^1)$-module ${\cal F}_{-1}$ ? The answer is: no, thanks to the triviality of
the cohomology space $H^2(\mathrm{Vect}(S^1),{\cal F}_{-1})$
(see \cite{Fuks86}, or \cite[chapter 4]{GuiRog05}).
In other words, any Lie algebra structure $\widetilde{[\ ,\ ]}$ on the vector
space $\mathrm{Vect}(S^1)\oplus{\cal F}_{-1}$ such that
\BEQ
\widetilde{[(X,\phi),(Y,\psi)]}=\left([X,Y]_{\mathrm{Vect}(S^1)},
{\rm ad}_{\mathrm{Vect}(S^1)}X.\psi-{\rm ad}_{\mathrm{Vect}(S^1)}Y.\phi
+B(X,Y)\right)
\EEQ
($B$ antisymmetric two-form on $\mathrm{Vect}(S^1)$)
is isomorphic to the Lie structure of $\cal W$.

So one may say that $\cal W$ and its central extensions are natural objects to look at.

\subsection{Some results on representations of ${\cal W}$}

We shall give in this subsection several results, the second of which
  certainly deserves deeper thoughts and will be developed in the future.

\begin{enumerate}
\item
\begin{proposition}('no-go theorem')
 There is no way to extend the usual representation of 
 $\mathfrak{alt}_1$ as conformal vector fields into
an embedding of $\cal W$ into the Lie algebra of vector fields on $\R^3$.
\end{proposition}

\noindent {\bf Proof:} 
Put $L_2^{\eps}=f\partial_t+g\partial_r+h\partial_{\zeta}$ 
where $f=f(t,r,\zeta),g=g(t,r,\zeta),h=h(t,r,\zeta)$ 
are yet undetermined functions. We use the explicit forms
of the generators of $\mathfrak{alt}_1$. Then the relations
\begin{displaymath}
{}[L_2^{\eps},L_{-1}]=3L_1^{\eps} \;\; , \;\;
{}[L_2^{\eps},L_{-1}^{\eps}]=0 \;\; , \;\;
{}[L_2^{\eps},L_0^{\eps}]=0
\end{displaymath}
give respectively
\BEA
\partial_r f &=& -6t^2,\ \partial_r g\:=\: -6tr,\ \partial_r h\:=\:
3\II r^2 \label{eq:one}  \\
\partial_{\zeta}f&=& \partial_{\zeta}g\:=\: \partial_{\zeta}h\:=\: 0\\
\partial_r f&=& 0,\  f\:=\: tr\partial_r g,\ \II g+t\partial_r h\:=\: 0.
\label{eq:three}
\EEA
But (\ref{eq:one}) and (\ref{eq:three}) are incompatible.
\hfill \qed 

\item
The following proposition hints at quite unexpected 
connections between contact structures in $\R^3$, 
the Lie algebra $\sv$ and the Lie algebra ${\cal W}$. 
Recall that a {contact form} $\al$ on a three-dimensional manifold 
${\cal V}$  is a one-form on $\cal V$ such that 
$\D \al \wedge \al$ is a non-degenerate volume form.
\begin{proposition}
Let $\alpha$ be the complex-valued contact form on $\R^3$
defined by $\al(t,r,\zeta)=r\D r-2\II \zeta \D t$.
Then the Lie algebra of vector fields $X(t,r,\zeta)$ such that:
\begin{enumerate}
\item [(i)] ${\cal L}_X\al$ is proportional to $\al$, 
i.e. ${\cal L}_X \al=f\al$ for a certain function $f=f(t,r,\zeta)$;
\item [(ii)] $[X,\partial_{\zeta}]=0$, i.e., components of $X$ do not depend on $\zeta$;
\item [(iii)] ${\cal L}_X \D t$ is proportional to $\D t$, so that $X$ is
tangent to each leaf $t=$constant
\end{enumerate}
is generated by $L_n,L_n^{\eps}$, $n\in \Z$, with
\BEA
L_n &=&-t^{n+1}\partial_t-\half(n+1)t^n r\partial_r
+{\II\over 4} (n+1)nt^{n-1}r^2 \partial_{\zeta} - \frac{x}{2}(n+1) t^n , \nonumber \\
L_n^{\eps} &=&-t^{n+1}{1\over r}\partial_r+{\II\over 2} (n+1)t^n \partial_{\zeta}.
\label{gl:Jrep}
\EEA
The Lie algebra $\langle L_n,L_n^{\eps}\rangle$ is isomorphic to 
$\cal W$, with commutators given by formula (\ref{gl:W}) in paragraph  2.3.
\end{proposition}

The attentive reader will have noted that  $L_n=X_n$ and
$L_n^{\eps}={1\over r}Y_{n+\half}$ \cite{Henk03}.

\item \begin{proposition} The infinite-dimensional extension ${\cal W}$
of the algebra $\mathfrak{alt}_1$ is a contraction of a pair
of commuting loop algebras
$\mathrm{Vect}(S^1)\oplus\overline{\mathrm{Vect}(S^1)}\rar{\cal W}$. In
particular, we have the explicit differential operator representation
\BEA
L_n &=& -t^{n+1}\partial_t - (n+1)t^n r\partial_r -(n+1)x t^n
-n(n+1)\gamma t^{n-1} r \nonumber \\
L_n^{\eps} &=& -t^{n+1}\partial_r -(n+1)\gamma t^n
\label{gl:formeexpl}
\EEA
where $x$ and $\gamma$ are parameters and $n\in\mathbb{Z}$.
\end{proposition}

\noindent {\bf Proof:} 
Let $\ell_n$ and $\bar{\ell}_n$ be the generators
of the two commuting loop algebras $\mathrm{Vect}(S^1)$ 
and $\overline{\mathrm{Vect}(S^1)}$. Obviously, the generators 
$X_n := \ell_n + \bar{\ell}_n$ and $Y_n := a \bar{\ell}_n$ satisfy the commutation relations
\begin{displaymath}
{}\left[ X_n, X_m\right] = (n-m) X_{n+m},
{}\left[ X_n, Y_m\right] = (n-m) Y_{n+m},
{}\left[ Y_n, Y_m\right] = a(n-m) Y_{n+m}
\end{displaymath}
which in the limit $a\to 0$ reduces to (\ref{gl:W}). A differential-operator
representation of the $X_n$ and $Y_n$ is given in case 
(iii) of table~1 of \cite{Henk02} and $L_n = \lim_{a\to 0} X_n$ and
$L_n^{\eps}=\lim_{a\to 0} Y_n$ which
yields the form (\ref{gl:formeexpl}).
\hfill \qed 

\end{enumerate}

One of the possible applications of these generators is the calculation of
multipoint correlation functions of many-body systems. One says that
the $n$-point correlator
$F_n := \left\langle \Phi_1\cdots \Phi_n\right\rangle$ of so-called
quasi-primary fields $\Phi_j$ is covariant under the
action of the generators $({\cal X}_i)_{i\in I}$ of a Lie algebra if
${\cal X}_i F_n=0$ for all $i\in I$. We apply this idea to the two-point
correlators covariant under $\mathfrak{alt}_1$. 
The standard representation (\ref{eq:zetaGen})
refers to the coordinates $\zeta,t,r$ and the quasi-primary field
will be denoted by $\psi=\psi(\zeta,t,r)$ and is assumed to have a scaling
dimension $x$. The two-point function reads \cite{Henk03}
\BEA 
\left\langle \psi_1(\zeta_1,t_1,r_1)\psi_2(\zeta_2,t_2,r_2)\right\rangle
 & = & (t_1-t_2)^{-(x_1+x_2)/2} \left(\frac{t_1}{t_2}\right)^{(x_2-x_1)/2}\times\nonumber\\
 & \times & f\left( \zeta_1-\zeta_2+\frac{\II}{2}\frac{(r_1-r_2)^2}{t_1-t_2}\right),\label{gl:phi_st}
\EEA
where $f$ is an arbitrary function. Similarly, imposing covariance under
the representation (\ref{gl:Jrep}) leads to
\BEQ \label{gl:phi_j}
\left\langle \psi_1(\zeta_1,t_1,r_1)\psi_2(\zeta_2,t_2,r_2)\right\rangle
= \delta_{x_1,x_2}\, (t_1-t_2)^{-x_1}
f\left( \zeta_1+\zeta_2+\frac{\II}{2}\frac{r_1^2-r_2^2}{t_1-t
_2}\right),
\EEQ
where again $f$ is an arbitrary function. It is evident that these two
representations will describe quite distinct physical systems.

Furthermore, instead to working with the variable $\zeta$, it is from a
physical point of view more natural to consider the Fourier-transform
of the field $\psi$ with respect to $\zeta$ and to define \cite{Henk03}
\BEQ
\phi(t,r) = \phi_{\cal M}(t,r) := \frac{1}{\sqrt{2\pi\,}}
\int_{\mathbb{R}} \!\D\zeta\, e^{-\II{\cal M}\zeta} \psi(\zeta,t,r)
\EEQ
Then the quasiprimary fields $\phi$ are characterized by $\cal M$ and their
scaling dimension. We find the following two-point correlation functions
\begin{enumerate}
\item for the standard representation  we find from (\ref{gl:phi_st})
\cite{Henk03}
\BEA
\left\langle \phi_1(t_1,r_1) \phi_2^*(t_2,r_2)\right\rangle & = &
\phi_0\, \delta({\cal M}_1-{\cal M}_2)
\Theta(t_1-t_2) \left(\frac{t_1}{t_2}\right)^{(x_2-x_1)/2}
\times\nonumber\\
& \times & (t_1-t_2)^{-(x_1+x_2)/2}\exp\left[ -\frac{{\cal M}_1}{2}\frac{(r_1-r_2)^2}{t_1-t_2}\right]\label{fixmtwopoint}
\EEA
where $\Theta(t)$ is the Heaviside function and $\phi_0$ a normalization
constant.
\item for the representation (\ref{gl:Jrep}) we find from (\ref{gl:phi_j})
\BEA
\left\langle \phi_1(t_1,r_1) \phi_2(t_2,r_2)\right\rangle & = &
\phi_0\, \delta_{x_1,x_2}\,\delta({\cal M}_1-{\cal M}_2) (t_1-t_2)^{-x_1}\times\nonumber\\
& \times & \exp\left[ -\frac{{\cal M}_1}{2}\frac{r_1^2-r_2^2}{t_1-t_2}\right]
\label{twopointJrep}
\EEA
\item finally, for the representation (\ref{gl:formeexpl}), the
quasiprimary field $\phi(t,r)$ is characterized by its scaling dimension
$x$ and the extra parameter $\gamma$. One has \cite{Henk02}
\BEQ
\left\langle \phi_1(t_1, r_1) \phi_2(t_2,r_2)\right\rangle =
\phi_0 \, \delta_{x_1,x_2}\,\delta_{\gamma_1,\gamma_2}\,
(t_1-t_2)^{-x_1} \exp\left[ - 2\gamma_1 \frac{r_1-r_2}{t_1-t_2}\right].
\EEQ
\end{enumerate}
Clearly, the form of these two-point correlators, notably their invariance
under time- or space-translations, are different.

\section{Casimir Operators and Matrix Representations for $\mathfrak{alt}_1$.}

{}From now on, we consider the finite-dimensional representations
of $\mathfrak{alt}_1$. We shall use the 
following notational conventions (see paragraph 2.1):
\begin{enumerate}
\item $\slin(2,\R)$ is spanned by $X_{\pm 1},X_0$.
\item The commutative algebra $h$ is spanned by $Y_{\pm 1},Y_0$.
\end{enumerate}
The non-zero commutators of $\alt$ are:
\BEQ
[X_n,X_m]=(n-m)X_{n+m}\;\;,\;\;[X_n,Y_m]=(n-m)Y_{n+m}\;\;;\;\;n,m \in \{\pm 1,0\}
\EEQ
In the notation of section~2,
$X_n = L_n$ and $Y_n = L_n^{\eps}$.

\subsection{Casimir operators}

The construction of such  operators is important, because in the vector-field
representation they correspond to the invariant differential operators.
Looking for second-order differential operator we write:
\BEQ
\hat S=a_{ij}X_iX_j+b_{ij}X_iY_j+c_{ij}Y_iY_j+d_iX_i+e_iY_i
\EEQ
Here the sum over repeated index is understood and $i,j \in \{\pm 1,0\}$.
{}From the conditions $[\hat S,X_i]=[\hat S,Y_i]=0$ the coefficients $a,b,c,d,e$
can be determined. The result is the following:
\BEA
\hat S &=& AS_0+S_1\;\;,\;\;
A=\mbox{\rm cste.} \nonumber \\
S_0 &=& X_{-1}Y_1+X_1Y_{-1}-2X_0Y_0 \nonumber \\
S_1 &=& Y_{-1}Y_1-Y_0^2
\EEA
The calculation for different representations gives the results:
\begin{enumerate}
\item In the physical representation (1.4)
\BEQ
\hat S = -t^2(2\II \partial_{\zeta}\partial_t+\partial_r^2)-\II t(2x-1)\partial_{\zeta}
\EEQ
which for the canonical scaling dimension of the wave function $x=1/2$ reduces
to the usual Schr\"odinger-operator in dynamical-mass representation, see
\cite{Henk03,Stoi05}.

\item In the representation  (\ref{gl:Jrep})
\BEQ
\hat S= \II A\,
\left(\frac{x}{2}-1\right)\partial_{\zeta}-\frac{1}{4}\partial_{\zeta}^2
\EEQ

The inverse Fourier transformation with respect to $\zeta$ 
of the wave function leads to a constant.
\item Finally for the representation (\ref{gl:formeexpl}) this
operator is again constant.
\end{enumerate}
This confirms the statement that the ``fixed mass'' 
or projective representation of $\mathfrak{alt}_1$
characterizes the wave function with its scaling dimension $x$
and the constant $\gamma $ instead of mass (the mass generator drops out).\footnote{If one admits values
$z\ne 1,2$ for the dynamical exponent, there are no representations of $\alt$ in terms of {\em local}
differential operators. However, for generic values of $z$, {\em non-local} representations in terms
for fractional differential operators can be constructed, which can be shown to close on the
solution space of an appropriate linear PDE of fractional order \cite{Henk11}.} 
\begin{remark}
In fact, when consider $d$ spacial dimensions 
that is a representation of $\mathfrak{alt}_d$ 
the compatibility with rotations require this constant 
$\gamma$ to be a vector \cite{Cherniha10} 
\BEQ 
\gamma\to \vec{\gamma}=(\gamma_1,...,\gamma_d).\nonumber
\EEQ
\end{remark}

\subsection{Matrix representations}

\begin{enumerate}
\item The adjoint representation can be obtained directly from
the commutators( $[\eta_i,\eta_j]=c_{ij}^k\eta_k $) in the following $6\times 6$-matrix form
\BEQ
Y_{-1}=c_{1j}^k=\left(\begin{array}{cccccc} 0& 0& 0& 0& -1& 0\\ 0& 0& 0& 0& 0& -2\\ 0& 0& 0& 0& 0& 0\\
0& 0& 0& 0& 0& 0\\ 0& 0& 0& 0& 0& 0\\ 0& 0& 0& 0& 0& 0\end{array}\right),
\ Y_{0}=c_{2j}^k=\left( \begin{array}{cccccc} 0& 0& 0& 1& 0& 0\\ 0& 0& 0& 0& 0& 0\\ 0& 0& 0& 0& 0& -1\\
0& 0& 0& 0& 0& 0\\ 0& 0& 0& 0& 0& 0\\ 0& 0& 0& 0& 0& 0\end{array}\right)
\EEQ

\BEQ
Y_1=c_{3j}^k=\left(\begin{array}{cccccc} 0& 0& 0& 0& 0& 0\\ 0& 0& 0& 0& 2& 0\\ 0& 0& 0& 0& 0& 1\\
0& 0& 0& 0& 0& 0\\ 0& 0& 0& 0& 0& 0\\ 0& 0& 0& 0& 0& 0\end{array}\right),
\ X_{-1}=c_{4j}^k=\left( \begin{array}{cccccc} 0& 0& 0& 0& 0& 0\\ 0& -1& 0& 0& 0& 0\\ 0& 0& -2& 0& 0& 0\\
0& 0& 0& 0& 0& 0\\ 0& 0& 0& 0& -1& 0\\ 0& 0& 0& 0& 0& -2\end{array}\right)
\EEQ
\BEQ
X_0=c_{5j}^k=\left(\begin{array}{cccccc} 1& 0& 0& 0& 0& 0\\ 0& 0& 0& 0& 0& 0\\ 0& 0& -1& 0& 0& 0\\
0& 0& 0& 1& 0& 0\\ 0& 0& 0& 0& 0& 0\\ 0& 0& 0& 0& 0& -1\end{array}\right),
\ X_1=c_{6j}^k=\left( \begin{array}{cccccc} 0& 2& 0& 0& 0& 0\\ 0& 0& 1& 0& 0& 0\\ 0& 0& 0& 0& 0& 0\\
0& 0& 0& 0& 2& 0\\ 0& 0& 0& 0& 0& 1\\ 0& 0& 0& 0& 0& 0\end{array}\right)
\EEQ

\item An embedding into $\mathfrak{su}(4)$ can be obtained by taking the 6-dimensional
restriction of the algebra $\mathcal{W}$ and writing the generators in its
Pauli-matrix form
\BEQ
Y_{-1}=\left(\begin{array}{cccc} 0& 0& 0& 0\\ 0& 0& -1& 0\\ 0& 0& 0& 0\\
0& 0& 0& 0\end{array}\right),
\ Y_{0}=1/2\left( \begin{array}{cccc} 0& 0& -1& 0\\ 0& 0& 0& 1\\ 0& 0& 0& 0\\
0& 0& 0& 0\end{array}\right)
\EEQ

\BEQ
Y_1=\left(\begin{array}{cccc} 0& 0& 0& 1\\ 0& 0& 0& 0\\ 0& 0& 0& 0\\
0& 0& 0& 0\end{array}\right),
\ X_{-1}=\left( \begin{array}{cccc} 0& 0& 0& 0\\ -1& 0& 0& 0\\ 0& 0& 0& 0\\
0& 0& -1& 0\end{array}\right)
\EEQ

\BEQ
X_0=1/2\left(\begin{array}{cccc} -1& 0& 0& 0\\ 0& 1& 0& 0\\ 0& 0& -1& 0\\
0& 0& 0& 1\end{array}\right),
\ X_1=\left( \begin{array}{cccc} 0& 1& 0& 0\\ 0& 0& 0& 0\\ 0& 0& 0& 1\\
0& 0& 0& 0\end{array}\right)
\EEQ
\end{enumerate}

\section{Cartan Decomposition and Dual Representations}

It is clear, from the commutation relations, that $\mathfrak{alt}_1$
has the following Cartan decomposition:\label{cartan}
\BEQ
\mathfrak{alt}_1={\cal P}\oplus {\cal K}\oplus {\cal L} = \{
Y_1,X_1\} \oplus \{ Y_0,X_0\} \oplus \{ Y_{-1},X_{-1}\}
\EEQ
and there is a one-to-one correspondence between the subalgebras ${\cal P}$
and ${\cal L}$.\\
The typical element ${\cal X} $ of the algebra $\mathfrak{alt}_1$ is given by
${\cal X}=\sum_{i=1}^6 \alpha_i\eta_i$ where 
$\{\eta_i\}, i=1,..,6$ is a basis of $\mathfrak{alt}_1$.
The $\{\alpha_i\}, i=1,\ldots,6$ are called {\em coordinates of the first kind}.
The matrix form is:
\BEQ
{\cal X}=\left(\begin{array}{cccc} -\alpha_4/2& \alpha_2&
-\alpha_3/2& \alpha_1\\
-\alpha _6& \alpha_4/2& -\alpha_5& \alpha_3/2\\0& 0& -\alpha_4/2& \alpha_2\\0&
0& -\alpha_6& -\alpha_4/2
\end{array}\right)
\EEQ
The group element (near to identity) can be expressed as:
\BEQ
\exp(\alpha_i\eta_i)=g(\{A_i \})=\exp(A_1\eta_1)\cdots \exp(A_6\eta_6)
\EEQ
The $A_i, i=1,\ldots,6$ are called {\em coordinates of the second kind}.
Here the following correspondence is made 
$\eta_1=Y_1, \eta_2=X_1, \eta_3=Y_0, \eta_4=X_0,\eta_5=Y_{-1}, \eta_6=X_{-1}$.
Next, consider the one-parameter subgroup generated by
${\cal X}, e^{s{\cal X}}$, the coordinates $\alpha $
scale by factor $s$, while the coordinates $A$ 
become functions of the single parameter $s$. Consequently one can write
\BEQ
g(A(s))=e^{s\mathcal{X}}.
\EEQ
Evaluating at $s=1$ gives the coordinate transformation $A=A(\alpha )$, 
while taking derivatives with respect to $A_i$ gives
\BEQ
{\cal X}g=\sum_ie^{A_1\eta_1}\cdots
e^{A_{i-1}\eta_{i-1}}\dot{A}_i\eta_ie^{A_i\eta_i}\cdots e^{A_6\eta_6}=
\dot{A}_{\mu}\partial_{\mu}g
\EEQ
with $\partial_{\mu}=\partial /\partial A_{\mu}$; 
the dot denotes differentiation with respect to $s$. 
Further considerations show that the coordinates $A$ 
contain the complete information about the Lie algebra structure.

We can calculate
\BEQ
g(\{ A_i\})=e^{-A_4/2}\left(\begin{array}{cccc} 1-A_2A_6e^{A_4}& A_2e^{A_4}& A& 
{\bar A}\\-A_6e^{A_4}& e^{A_4}& -(A_5+\frac{1}{2}A_3A_6)e^{A_4}& \frac{1}{2}A_3e^{A_4}\\
0& 0& 1-A_2A_6e^{A_4}& A_2e^{A_4}\\
0& 0& -A_6e^{A_4}& e^{A_4}\end{array}\right)\label{groupstructure}
\EEQ
where \BEQ A = -(A_2A_5+\frac{1}{2}{A_2A_3A_6}+A_1A_6)e^{A_4}-\frac{1}{2}{A_3}, \quad
{\bar A}= (\frac{1}{2}{A_2A_3}+A_1)e^{A_4}.\nonumber\EEQ
From (\ref{groupstructure}) the second kind coordinates 
can be given in terms of the elements
of the matrix representation of the group (and conversely):
\BEA
A_1=g_{14}/g_{22}-g_{12}g_{24}/(g_{22}^2)\;\;,\;\;
A_2=g_{12}/g_{22}\;\;,\;\;
A_3=2g_{24}/g_{22} \nonumber \\
A_4=2\ln g_{22} \;\;,\;\;
A_5=-g_{23}/g_{22}-g_{24}g_{21}/(g_{22}^2) \;\;,\;\;
A_6=-g_{21}/g_{22}.
\EEA
The multiplication by basis elements $\eta , g\mapsto g\eta $, acting on the universal enveloping
algebra with basis $[n]=\eta^n=\eta_1^{n_1}....\eta_6^{n_6}$ are realized as left-invariant vector
fields $\eta^*$, acting on function of $A$ (action commutes with multiplication by group element
on the left, so $\eta $ acts on the right), given in terms of pi-matrix 
$\eta^*_i=\pi^*_{i\mu}(A)\partial_{\mu}$.
Similarly, multiplication on the left gives right-invariant vector fields
$\eta_i^{\ddag}=\pi_{i\mu }^{\ddag}(A)\partial_{\mu }$.

The dual representations are defined as realization of the Lie algebra
as vector fields in terms of coordinates of 
the second kind acting on the left or right respectively
\BEQ
\eta_jg(A)=\pi_{j\mu}^{\ddag}(A)\partial_{\mu}g(A)\;\;,\;\;
g(A)\eta_j=\pi^*_{j\mu}(A)\partial_{\mu}g(A)\label{duallemma}.
\EEQ
The connection between left and right dual representations 
is given by the following splitting lemma\\
\begin{lemma}
\BEQ
\dot{A}_k=\alpha_{\mu}\pi_{\mu k}^*(A)=\alpha_{\mu}\pi_{\mu k}^{\ddag}(A)
\EEQ
with initial values $A_k(0)=0,\pi^*(0)=\pi^{\ddag}(0)=I$, is gathered from 
\cite{AVR,Fein93a,Fein93c}.
\end{lemma}
For our case we find
\BEQ
\pi^{\ddag}=\left(\begin{array}{cccccc} 1& 0& 0& 0& 0& 0\\ 0& 1& 0& 0& 0& 0\\
0& 0& 1& 0& 0& 0\\ -A_1& -A_2& 0& 1& 0& 0\\ A_2^2& 0& -2A_2& 0& e^{-A_4}& 0\\
2A_1A_2& A_2^2& -2A_1& -2A_2& -A_3e^{-A_4}& e^{-A_4}\end{array}\right)
\EEQ
\BEQ
\eta^{\ddag}=\left( \begin{array}{c} \partial_1\\ \partial_2\\ \partial_3\\
-A_1\partial_1-A_2\partial_2-\partial_4\\
A_2^2\partial_1-2A_2\partial_3+e^{-A_4}\partial_4\\
2A_1A_2\partial_1+A_2^2\partial_2-2A_1\partial_3-2A_2\partial_4-
A_3e^{-A_4}\partial_5-e^{-A_4}\partial_6\end{array}\right)
\EEQ

\BEQ
\pi^*=\left(\begin{array}{cccccc} e^{-A_4}& 0& -2A_6& 0& A_6^2& 0\\
-A_3e^{-A_4}& e^{-A_4}& -2A_5& -2A_6& 2A_5A_6& A_6^2\\
0& 0& 1& 0& -A_6& 0\\ 0& 0& 0& 1& -A_5& -A_6\\
0& 0& 0& 0& 1& 0\\ 0& 0& 0& 0& 0& 1\end{array}\right)
\EEQ
\BEQ
\eta^*=\left( \begin{array}{c} e^{-A_4}\partial_1-2A_6\partial_3+A_6^2\partial_5\\
-A_3e^{-A_4}\partial_1+e^{-A_4}\partial_2-2A_5\partial_3-2A_6\partial_4+2A_5A_6\partial_5+A_6^2\partial_6\\
\partial_3-A_6\partial_5\\
\partial_4-A_5\partial_5-A_6\partial_6\\
\partial_5\\
\partial_6\end{array}\right)
\EEQ

The last representation leads to the physical case (\ref{gl:formeexpl}) 
from \cite{Henk02} if $A_5\mapsto -r, A_6\mapsto -t$
and one supposes the action of the vector fields on the 
functions in the form  $e^{-\gamma A_3}e^{-xA_4}f(A_5, A_6)$.

\section{Wick Products and Appell Polynomials}

{\em Appell polynomials} share many properties with Wick products.
In physical literature, the term Wick product is even more popular.
The aim of this section is to provide a ``smooth'' 
introduction to Appell polynomials through Wick products.
The following presentation is gathered from \cite{AVR}.

Let $X_1, X_2,\ldots$ be random variables. The Wick powers are defined
inductively on $k$ as follows.
Start with $\langle X_1,X_2,\ldots,X_k\rangle =1$ for $k=0$.
Then for any $k>0$,  $\langle X_1,X_2,\ldots,X_k\rangle $
is defined recursively for $k=1,2,\ldots$, by
\BEQ
           E\langle X_1,X_2,\ldots,X_k\rangle=0
\EEQ
and
\BEQ
    \frac{\partial \langle X_1,X_2,\ldots,X_k \rangle }{\partial X_i}
      = \langle X_1,\ldots,X_{i-1},{\hat X}_i, X_{i+1},\ldots,X_k \rangle
\EEQ
where E means expectation (or mean) and ${\hat X}_i$ denotes the absence of the $X_i$ variable.\\
\\
\begin{example}
The first two Wick products are
\begin{eqnarray}
       <X_1>&=&X_1-EX_1 \nonumber \\
   <X_1,X_2>&=&X_1X_2-X_1EX_2-X_2EX_1+2EX_1EX_2-EX_1X_2
\end{eqnarray}
The Appell polynomials $P_n(x)$ are then defined by
\BEQ
          P_{X,n}(X)=P_n(X)=\underbrace{<X,\ldots,X>}_{n\rm\;times}
\EEQ
\end{example}
\begin{example}
Denoting $m_1=EX=0$ and $m_i=EX^i, i=2,\ldots$, we have:
\begin{eqnarray*}
 P_0(x)&=&1 \\
 P_1(x)&=&x \\
 P_2(x)&=&x^2-m_2 \\
 P_3(x)&=&x^3-m_3-3m_2x \\
 P_4(x)&=&x^4-10m_2x^3-10m_3x^2+5x(6m^2_2-m_4) \\
 P_5(x)&=&x^5-10m_2x^3-10m_3x^2+5x(6m^2_2-m_4)+20m_2m_3-m_5
\end{eqnarray*}
\end{example}
\begin{remark}
If $X\sim N(0,1)$ (the gaussian random variable with mean equal to 
$0$ and variance equal to $1$), then we get the familiar Hermite polynomials.
But in general, Appell polynomials are not necessarily orthogonal polynomials.
\end{remark}
Appell polynomials $P_n(x);n\in\mathbb{N}$
are also characterized by the two conditions
\begin{enumerate}
\item $P_n(x)$ is a polynomial of degree $n$,
\item $\frac{\D}{\D x}P_n(x)=nP_{n-1}(x)$
\end{enumerate}
Interesting examples are furnished by the shifted moment sequence
\BEQ P_n(x)=\int^{\infty}_{-\infty}(x+y)^n\mu(\D y),\label{Appellprob}\EEQ
where $\mu$ is a probability measure on $\R$ with all moments finite.
Of course, this includes in particular the Hermite polynomials for the
Gaussian case.
In \cite{FS1} the probabilistic interpretation of Appell polynomials is used
to define their analog on Lie groups where, in general, they are no longer
polynomials. For this reason they are called {\em Appell systems}.

\section{Appell Systems of the Algebra $\mathfrak{alt}_1$}

Appell systems of the Schr\"odinger algebra $\mathfrak{sch}_1$ have been
investigated in \cite{Fein04} but the algebra $\mathfrak{alt}_1$ requires a
specific study.\\
Referring to the decomposition (\ref{cartan}), we specialize variables,
writing $V_1,V_2,B_1,B_2$ for $A_1,A_2,A_5,A_6$ respectively.
Basic for our approach is to calculate
$e^{B_1Y_{-1}+B_2X_{-1}}e^{V_1Y_1+V_2X_1}$. We get
\begin{equation}
B_1Y_{-1}+B_2X_{-1}=\left(\begin{array}{cccc} 
0& 0& 0& 0\\ 
-B_2& 0& -B_1& 0\\ 
0& 0& 0& 0\\ 
0& 0& -B_2& 0
\end{array}\right)
\;\; , \;\;
V_1Y_1+V_2X_1=\left(\begin{array}{cccc} 
0& V_2& 0& V_1\\ 
0& 0& 0& 0\\ 
0& 0& 0& V_2\\ 
0& 0& 0& 0
\end{array}\right)
\end{equation}
and finally:
\begin{equation}
e^{B_1Y_{-1}+B_2X_{-1}}e^{V_1Y_1+V_2X_1}=
\left(\begin{array}{cccc} 1& V_2& 0& V_1\\ -B_2& 1-B_2V_2& -B_1& -B_2V_1-B_1V_2\\
0& 0& 1& V_2\\ 0& 0& -B_2& 1-B_2V_2\end{array}\right)
\end{equation}
\begin{proposition} In coordinates of the second kind, we have the Leibniz formula
\begin{eqnarray}\label{eq:pglaw}
&& g(0,0,0,0,B_1,B_2)g(V_1,V_2,0,0,0,0)= g(A_1,A_2,A_3,A_4,A_5,A_6)=
\nonumber \\
&=& g\left({B_1V_2^2+V_1\over (1-B_2V_2)}, {V_2\over (1-B_2V_2)}, -2{B_1V_2+B_2V_1\over (1-B_2V_2)},\right.\nonumber \\
&&\left. ~~ \ln (1-B_2V_2), {B_1-2B_1B_2V_2-B_2^2V_1\over (1-B_2V_2)^2}, {B_2\over (1-B_2V_2)}\right)
\end{eqnarray}
\end{proposition}
Now we are ready to construct the representation space and basis --
the canonical Appell system. To start,
define a vacuum state $\Omega$. The elements $Y_1,X_1$ of $\mathfrak{P}$
can be used to form basis elements
\begin{equation}
|jk\rangle =Y_1^jX_1^k\Omega ,j,k\geqslant 0
\end{equation}
of a Fock space $\mathfrak{F}=\mathrm{span}\{|jk\rangle \}$ on which $Y_1,X_1$ act as raising
operators,$Y_{-1},X_{-1}$ as lowering operator and $Y_0,X_0$ as multiplication
with the constants $\gamma,x $ (up to the sign) correspondingly. That is,
\begin{eqnarray}
&&Y_1\Omega =|10\rangle , X_1\Omega=|01\rangle \nonumber \\
&&Y_{-1}\Omega=0, X_{-1}\Omega=0 \label{eq:fock} \\
&&Y_0\Omega=-\gamma |00\rangle , X_0\Omega = -x|00\rangle \nonumber
\end{eqnarray}
The goal is to find an abelian subalgebra spanned by some self-adjoint
operators acting on the representation space, just constructed. Such a
two-dimensional subalgebra can be obtained by an appropriate ``turn'' of
the plane ${\cal P}$ in the Lie algebra, namely via the adjoint action of
the group element formed by exponentiating $X_{-1}$. The resulting plane,
${\cal P}_{\beta}$ say, is abelian and is spanned by
\begin{eqnarray}
\bar{Y_1} &=& e^{\beta X_{-1}}Y_1e^{-\beta X_{-1}}=Y_1-2\beta Y_0 +\beta^2Y_{-1} \nonumber \\
\bar{X_1} &=& e^{\beta X_{-1}}X_1e^{-\beta X_{-1}}=X_1-2\beta X_0 +\beta^2X_{-1} \label{eq:abepl}
\end{eqnarray}
Next we determine our canonical Appell systems. 
We apply the Leibniz formula (\ref{eq:pglaw}) with
$B_1=0$, $B_2=\beta ,V_1=z_1, V_2=z_2$ and use (\ref{eq:fock}). This yields
\begin{eqnarray}\label{eq:genfu}
e^{z_1\bar{Y_1}}e^{z_2\bar{X_1}}\Omega &=& 
e^{\beta X_{-1}}e^{z_1Y_1}e^{z_2X_1}e^{-\beta X_{-1}}\Omega=
e^{\beta X_{-1}}e^{z_1Y_1}e^{z_2X_1}\Omega \nonumber \\
&=& e^{z_1Y_1\over (1-\beta z_2)^2}e^{z_2X_1\over (1-\beta z_2)}
e^{2\gamma\beta z_1\over (1-\beta z_2)}(1-\beta z_2)^{-2x}\Omega
\end{eqnarray}
To get the generating function for the basis $|jk\rangle$ set in equation (\ref{eq:genfu})
\begin{equation}\label{eq:chang}
v_1={z_1\over (1-\beta z_2)^2},\quad v_2={z_2\over (1-\beta z_2)}
\end{equation}
Substituting throughout, we have
\begin{proposition}
The generating function for the canonical Appell system\\ $|jk\rangle =Y_1^jX_1^k\Omega$ is:
\begin{eqnarray}\label{eq:appsy}
e^{v_1Y_1+v_2X_1}\Omega
&=& \exp\left(y_1{v_1\over (1+\beta v_2)^2}\right)
\exp\left(y_2{v_2\over (1+\beta v_2)}\right)
\exp\left(-{2\gamma \beta v_1\over (1-\beta v_2)}\right)\nonumber \\
& \times & (1+\beta v_2)^{-2x}\Omega,\label{genfun}
\end{eqnarray}
where we identify $\bar{Y_1}\Omega =y_1 \cdot 1$ and
$\bar{X_1}\Omega =y_2 \cdot 1$
in the realization as function of $y_1,y_2$.
\end{proposition}
\begin{remark}
With $v_1=0$, we recognize the generating function for the Laguerre polynomials,
while $v_2=0$ reduces to the generating function of a standard Appell system.
\end{remark}
\section{Coherent States and Leibniz Function}

Now we define an inner product such that
\BEQ\label{eq:opcon}
 Y_1^{\dag}=\beta^2Y_{-1} \;\;,\;\; X_1^{\dag}=\beta^2X_{-1}.
\EEQ
In such a way the operators (\ref{eq:abepl}) are extended to self-adjoint ones
on appropriate domains. For simplicity
we take $\beta=1$ and define the two-parameter family of {\em coherent states}.
\BEQ\label{eq:cohst}
\Psi_V=\Psi_{V_1,V_2}=e^{V_1Y_1}e^{V_2X_1}\Omega .
\EEQ
The Leibniz function is defined as inner product of coherent states
\BEA
\mathcal{Y}_{BV} &=& \left\langle \Psi_B, \Psi_V\right\rangle = \left\langle
\Omega,e^{B_1Y_{-1}} e^{B_2X_{-1}}
e^{V_1Y_1} e^{V_2X_1}\Omega \right\rangle = \nonumber \\
&=& (1-B_2V_2)^{-2x}\exp\left( {2\gamma (B_1V_2+B_2V_1)\over (1-B_2V_2)}\right)
\label{leibf}
\EEA
Here we use the result (\ref{eq:pglaw}) and a normalization
$\left\langle\Omega,\Omega\right\rangle =1$
is understood. Further consideration shows that one can recover the raising and
lowering operators as elements of the Lie algebra acting on the Hilbert space with basis consisting of the
canonical Appell systems.

The remarkable fact is that the Lie algebra can be reconstructed from the
Leibniz function $\mathcal{Y}_{BV}$.
Really, differentiation with respect to $V_1$ brings down $Y_1$ acting on
$\Psi_V$, while differentiation
with respect to $B_1$ bring down $Y_1$ acting on $\Psi_B$ which moves across
the inner product as $Y_{-1}$
acting on $\Psi_V$. Similarly for $X_1$ and $X_{-1}$. We thus introduce creation
operators ${\cal R}_i$
and annihilation operators ${\cal V}_i$, satisfying
$[{\cal V}_i,{\cal R}_i]=\delta_{ig}I$.
For $\mathfrak{alt}_1$ we identify $Y_1={\cal R}_1, X_1={\cal R}_2$.
Note however, that ${\cal V}_1$
is not adjoint of ${\cal R}_1$, nor ${\cal V}_2$ of ${\cal R}_2$.
Bosonic realization of the respective
adjoints $Y_{-1}, X_{-1}$, we want to determine now. One method is the
following. When the explicit form of the
Leibniz function $\mathcal{Y}_{BV}= \mathcal{Y}$ is known, one can (formally)
write the partial differential equations for it. In our case they are
\BEA
\partial_{B_1}\mathcal{Y} &=& (V_2^2\partial_{V_1}+2\gamma V_2)\mathcal{Y}
\nonumber \\
\partial_{B_2}\mathcal{Y} &=&
(V_2^2\partial_{V_2}+V_1V_2\partial_{V_1}+2xV_2+2\gamma V_1)
\mathcal{Y} \label{eq:detlf}
\EEA
Then, one interprets each multiplication by $V_i$ as the operator
${\cal V}_i$ and each differentiation by $V_i$ as the operator ${\cal R}_i$. This gives the
following action of the operators $Y_{-1},X_{-1}$ on polynomial functions of $Y_1$ and $X_1$:
\BEA
Y_{-1} &=& 2\gamma {\cal V}_2+{\cal R}_1{\cal V}_2^2 \nonumber \\
X_{-1} &=&
{\cal R}_2{\cal V}_2^2+2{\cal R}_1{\cal V}_1{\cal V}_2
+2\gamma {\cal V}_1+ 2x{\cal V}_2. \label{eq:imres}
\EEA
{}From the commutation relations we find $Y_0$ and $X_0$
\BEQ
Y_0=-{\cal R}_1{\cal V}_2-\gamma ,
X_0=-{\cal R}_1{\cal V}_1-{\cal R}_2{\cal V}_2-x
\EEQ
 Using the more usual notation $a_1={\cal V}_1, a_2={\cal V}_2,
a_1^+={\cal R}_1, a_2^+={\cal R}_2$ we have the following:
\begin{proposition}
The raising $a_1^+, a_2^+$ and lowering $a_1, a_2$ operators appear as elements of the algebra
$\mathfrak{alt}_1$ in the following way:
\BEA
 Y_1 &=& a_1^+, \quad X_1 = a_2^+ \nonumber \\
 Y_0 &=& -a_1^+a_2-\gamma, \quad X_0 = -a_1^+a_1-a_2^+a_2-x\nonumber\\
 Y_{-1} &=& a_1^+a_2^2+2\gamma a_2, \quad X_{-1} = a_2^+a_1^2+2a_1^+a_1a_2+2\gamma a_1+2xa_2 \label{eq:bosre}
\EEA
The action on the Fock space, consisting of two parameter family of coherent state(\ref{eq:cohst})
with basis the canonical Appell systems with generated function (\ref{genfun}) is as follows:
\BEA
 Y_1|j,k\rangle &=& |j+1,k\rangle, \quad X_1|j,k\rangle = |j,k+1\rangle \nonumber \\
 Y_0|j,k\rangle &=& -k|j+1,k-1\rangle -\gamma |j,k\rangle, 
 \quad X_0|j,k\rangle = -(j+k+x)|j,k\rangle \nonumber \\
 Y_{-1}|j,k\rangle &=& k(k-1)|j+1,k-1\rangle +2\gamma |j,k-1\rangle\nonumber\\
 X_{-1}|j,k\rangle &=& k(k+2j+2x-1)|j,k-1\rangle+ 2j\gamma |j-1,k\rangle
\label{eq:acti}
\EEA
\end{proposition}
The natural involution on the algebra $Y_1 \leftrightarrow Y_{-1},
X_1\leftrightarrow X_{-1}$,
with substitutions $-a_1\leftarrow r, -a_2\leftarrow t, -a_1^+\leftarrow
\partial_r, -a_2^+\leftarrow \partial_t$ leads to the physical
representation (\ref{gl:formeexpl}).

\section{Concluding Remarks}

We have studied properties of the Poincar\'e or `altern' algebra $\mathfrak{alt}_1$, nowadays also
often referred to as `conformal galilean algebra', 
and of interest in connection with the ageing phenomenon in condensed-matter physics 
and in string-theory in the context of the AdS/CFT correspondence. In particular, 
we have shown that $\mathfrak{alt}_1$ can be embedded in an infinite-dimensional 
Lie algebra and have discussed its relationship with the Virasoro and the
Schr\"odinger-Virasoro algebras.\\
As for the representation-theory, we have systematically
constructed the Casimir operators and have written down explicit matrix
representations. Considerations of the Wick product has led us to the
construction of the Appell systems of $\mathfrak{alt}_1$ which are useful for the
construction of coherent states.\\
A more general study of random walks and stochastic processes on 
$\mathfrak{alt}_1$ is a challenging research project.

\section*{Acknowledgments}
The authors have been supported by the EU Research Training Network HPRN-CT-2002-00279.


\end{document}